\documentclass[conference]{IEEEtran}
\IEEEoverridecommandlockouts
\usepackage{cite}
\usepackage{amsmath,amssymb,amsfonts}
\usepackage{algorithmic}
\usepackage{graphicx}
\usepackage{textcomp}
\usepackage{xcolor}
\usepackage{url}

\def\BibTeX{{\rm B\kern-.05em{\sc i\kern-.025em b}\kern-.08em
    T\kern-.1667em\lower.7ex\hbox{E}\kern-.125emX}}
    
\begin{document}

\title{Tensor Network Circuit Simulation at Exascale
\thanks{Horizon 2020 Research and Innovation programme (2014-2020) under grant agreement 823767}
}

%

\DeclareRobustCommand*{\IEEEauthorrefmarknum}[1]{%
  \raisebox{0pt}[0pt][0pt]{\textsuperscript{\footnotesize #1}}%
}





\author{
\IEEEauthorblockN{1\textsuperscript{st} John Brennan}
\IEEEauthorblockA{\textit{Irish Centre for} \\
\textit{High End Computing} \\
Dublin, Ireland \\
john.brennan@ichec.ie}

\and

\IEEEauthorblockN{2\textsuperscript{nd} Momme Allalen}
\IEEEauthorblockA{\textit{Leibniz Supercomputing Centre of} \\
\textit{the Bavarian Academy of}\\
\textit{Sciences and Humanities}\\
Garching b. München, Germany \\
momme.allalen@lrz.de}

\and

\IEEEauthorblockN{3\textsuperscript{rd} David Brayford}
\IEEEauthorblockA{\textit{Leibniz Supercomputing Centre of} \\
\textit{the Bavarian Academy of}\\
\textit{Sciences and Humanities}\\
Garching b. München, Germany \\
david.brayford@lrz.de}

\and

\IEEEauthorblockN{4\textsuperscript{th} Kenneth Hanley}
\IEEEauthorblockA{\textit{Irish Centre for} \\
\textit{High End Computing} \\
Dublin, Ireland}

\and

\IEEEauthorblockN{5\textsuperscript{th} Luigi Iapichino}
\IEEEauthorblockA{\textit{Leibniz Supercomputing Centre of} \\
\textit{the Bavarian Academy of}\\
\textit{Sciences and Humanities}\\
Garching b. München, Germany \\
luigi.iapichino@lrz.de \\
ORCID:0000-0003-3938-8973}

\and

\IEEEauthorblockN{6\textsuperscript{th} Lee J. O'Riordan}
\IEEEauthorblockA{\textit{Irish Centre for} \\
\textit{High End Computing} \\
Dublin, Ireland \\
ORCID:0000-0002-6758-9433}

\and

\IEEEauthorblockN{7\textsuperscript{th} Myles Doyle}
\IEEEauthorblockA{\textit{Irish Centre for} \\
\textit{High End Computing} \\
Dublin, Ireland}

\and

\IEEEauthorblockN{8\textsuperscript{th} Niall Moran}
\IEEEauthorblockA{\textit{Irish Centre for} \\
\textit{High End Computing} \\
Dublin, Ireland \\
ORCID:0000-0002-2619-5040}
}

\maketitle

\begin{abstract}
Tensor network methods are incredibly effective for simulating quantum circuits. This is due to their ability to efficiently represent and manipulate the wave-functions of large interacting quantum systems.
We describe the challenges faced when scaling tensor network simulation approaches to Exascale compute platforms and introduce QuantEx, a framework for tensor network circuit simulation at Exascale.
\end{abstract}

\begin{IEEEkeywords}
quantum circuit, tensor network, exascale, HPC
\end{IEEEkeywords}

\section{Introduction}
The ability to simulate quantum circuits is essential for the design and development of quantum computing hardware and algorithms. A common method to simulate quantum circuits with a small number of qubits is to store and evolve the state vector of the circuit's qubits~\cite{Guerreschi_2020, Jones_2018, Smelyanskiy_2016}. However, for circuits with a larger number of qubits, such as those considered for Noisy Intermediate Scale Quantum (NISQ) devices~\cite{Preskill_2018, Brooks_2019}, it can be intractable to directly evolve the full quantum wave-function, even on the largest supercomputers~\cite{villalonga_qsim_2020}. This is due to the memory requirements to store the wave function which grows exponentially with the number of qubits in the circuit.

Alternative simulation methods for quantum circuits, which attempt to mitigate the exponential memory issue, are those based on tensor networks. With these methods, quantum circuits are represented as networks of tensors, enabling output probability amplitudes to be calculated by contracting the network~\cite{Bridgeman_2017, biamonte2017tensor}. This approach has achieved state of the art performance when simulating Random Quantum Circuits (RQC)~\cite{Boixo_2018} as part of the recent quantum advantage experiments~\cite{villalonga_qsim_2020}.

\begin{figure}[!t]
\centering
\includegraphics[width=.4\textwidth]{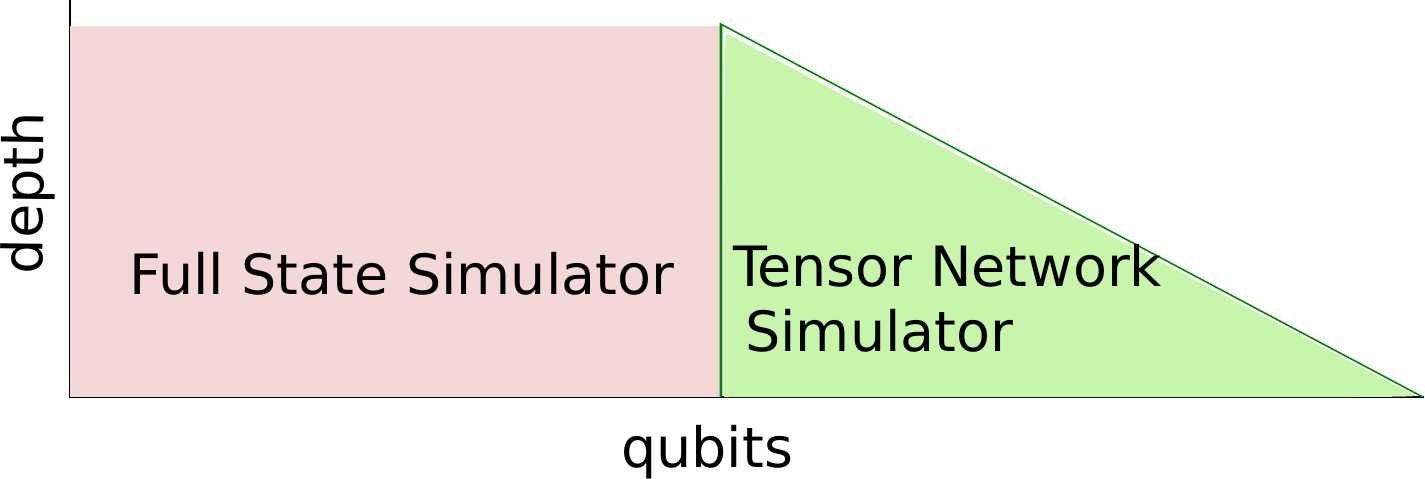}
\caption{The applicability of tensor network approaches is illustrated in the above cartoon. For quantum circuits with a small number of qubits, such that the full quantum state of the qubits can be stored in memory, full state simulators are expected to be the most performant. Tensor network based simulators can be used to simulate circuits with a larger number of qubits, where a full state approach is infeasible due to memory requirements. However, the maximum depth of a circuit which can be simulated using tensor networks decreases as the number of qubits in the circuit increases.}
\label{fig:area_of_application}
\end{figure}

Despite these impressive results, tensor network methods are not competitive for simulating all circuit types. In particular, for very deep/highly entangled circuits, the tensor network representation can require the same amount of memory as full wave-function methods. In these cases, full wave-function approaches with simpler memory management and fewer overheads are generally more efficient. For circuits targeting NISQ devices, with moderate depth/entanglement and where approximate results suffice, tensor network approaches can offer significant advantages as illustrated in Fig.~\ref{fig:area_of_application}.

The aim of this paper is to describe some of the challenges faced in utilising distributed platforms to simulate quantum circuits with tensor network methods and to present QuantEx, an open source quantum circuit simulation framework designed to be scalable and extensible. The paper is structured as follows: In Section \ref{sec:overview} we provide an overview of the main tasks of a tensor network simulator and outline how tensor networks can be used to simulate a quantum circuit. In Section~\ref{sec:scaling_challenges} we describe the challenges encountered when implementing an efficient tensor network simulator capable of leveraging distributed compute resources. We also provide some information on some of the commonly used methods for dealing with these challenges. We then present our own tensor network based quantum circuit simulator, QuantEx, in Section~\ref{sec:quantex} before concluding with an outlook on future work for the QuantEx project.

\section{Overview of a Tensor Network Simulator}\label{sec:overview}
Quantum circuit simulators are used to perform two important tasks. Namely, validating the correctness of quantum hardware and simulating the execution of a quantum algorithm as if it was executed on a real quantum device. Depending on which task the simulator is being used to do, the output of a simulator can either be a list of probability amplitudes, corresponding to a predefined list of possible output bitstrings, or a set of random bitstrings, which are distributed according to the output wavefunction of the simulated circuit. For instance, to validate the output from Google's Sycamore circuit, the authors of~\cite{Arute_2019} had to compute the probability amplitudes for a list of bitstrings which were output from their circuit. To sample random bitstrings from a circuit, a version of rejection sampling is often used~\cite{qflex_npj_2019, Markov_2018, Pan_2021} where a candidate bitstring $x$ is generated, usually with a uniform distribution, and is randomly accepted as a sample from the circuit with a probability proportional to the probability the circuit would output $x$ when measured. To do this, a probability amplitude for $x$ needs to be computed. These examples highlight the main job of a quantum circuit simulator which is to compute probability amplitudes for measurement outcomes of a predefined quantum circuit.

In order to produce probability amplitudes for different measurement outcomes, a simulator needs to store a representation of the simulated circuit's output state or wave function. Full state simulators do this by storing in memory a vector, representing the initial state of the circuit's qubits, and applying a sequence of unitary matrices, representing quantum gates in the circuit, to transform the initial state into the final state of the circuit's qubits. The desired probability amplitudes can then be read from this final vector. As mentioned in the introduction, a drawback of this method for larger circuits is that the size of the vector that needs to be stored grows exponentially with the number of qubits in the circuit. More precisely, a vector representing the state of $n$ qubits in a circuit needs to store $2^n$ probability amplitudes corresponding to the $2^n$ possible measurement outcomes. For large circuits, where the size of circuit's state vector exceeds the available memory resources, a tensor network representation of the circuit's output state can be used and is straight forward to create as we describe below after briefly introducing tensor networks.

\begin{figure}[!t]
\centering
\includegraphics[width=.48\textwidth]{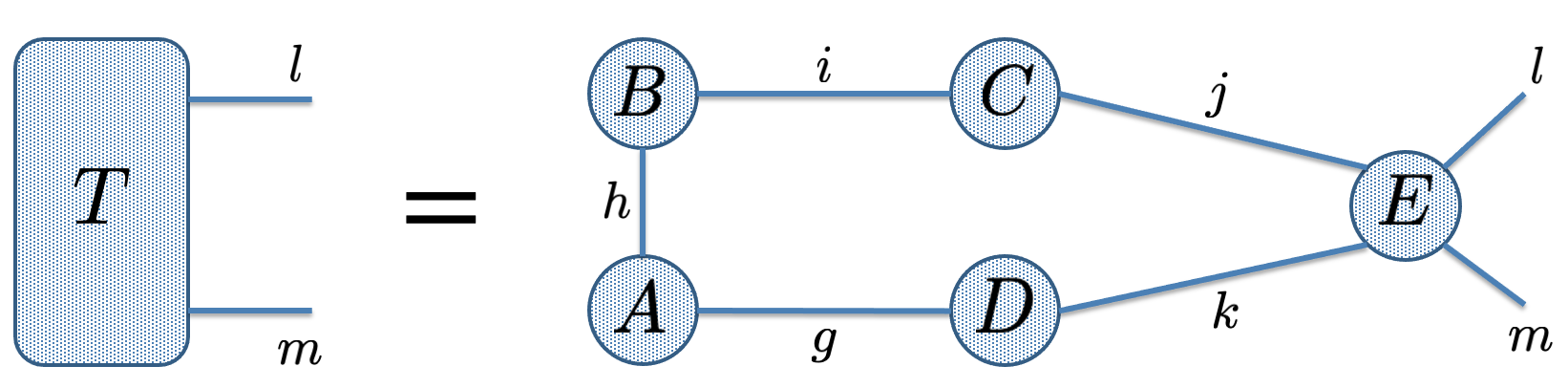}
\caption{A tensor network example.}
\label{fig:TN_expression}
\end{figure}

A tensor network is a tensor expression involving the product of several tensors, or multi-dimensional arrays of complex numbers, of varying rank or dimension. For example, in the following expression the right hand side is an example of a tensor network:
\begin{equation}\label{tn_expression}
    T_{lm} = \sum_{ghijk}A_{gh}B_{hi}C_{ij}D_{gk}E_{jklm}
\end{equation}
However, tensor networks typically refer to such expressions involving a large number of tensors such that a graphical notation is often used in place of the more traditional notation used above~\cite{Bridgeman_2017}. The graphical notation can be summarized as follows: Tensors in the tensor expression are depicted by nodes in a graph and indices in the expression are depicted by edges in the a graph. If two tensors share a common index then both tensors are connected by the edge representing that index. Indices that are unique to a tensor are depicted by an open edge. In graphical notation, equation~\eqref{tn_expression} can be written as shown in figure Fig.~\ref{fig:TN_expression}.

\begin{figure}[!t]
\centering
\includegraphics[width=.48\textwidth]{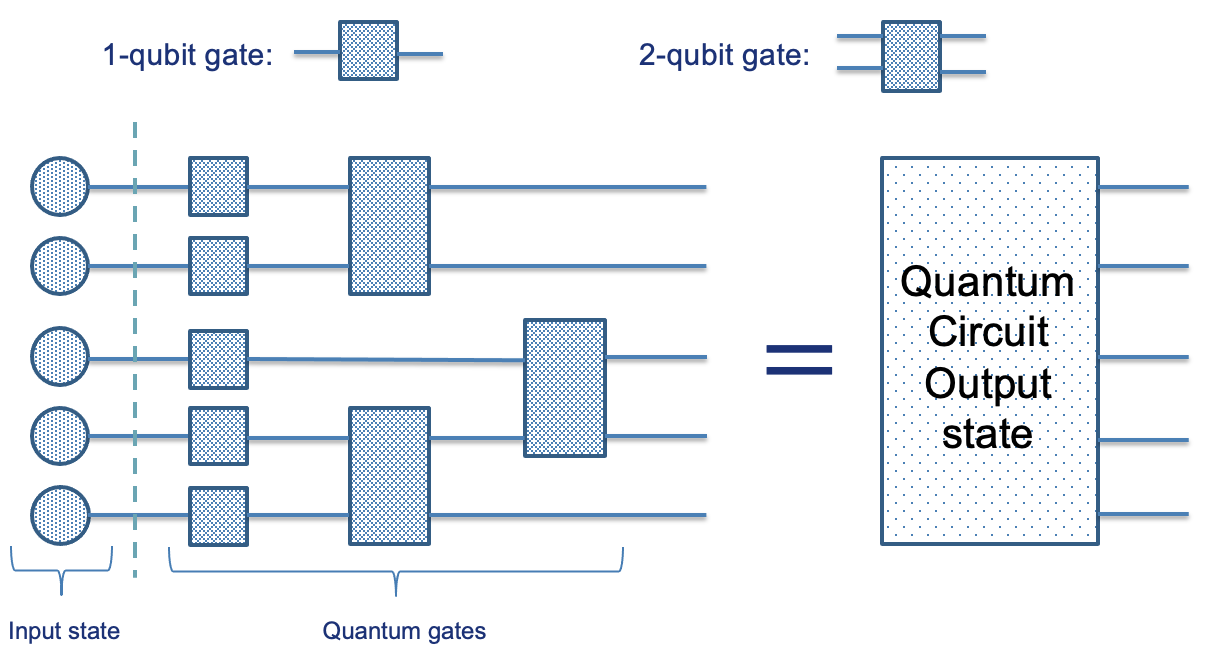}
\caption{An example of a tensor network representing the output quantum state a quantum circuit. The initial state of the circuit's qubits is encoded in the tensors appearing on the left hand side of the network. One and two qubit gates acting on the qubits are represented by tensors of rank two and four respectively. Contracting the tensor network results in a single large tensor containing the full state representation of the circuit's output state.}
\label{fig:QC_TN_expression}
\end{figure}

Given a complete description of a quantum circuit, a tensor network representing the output state of the circuit can be created using the matrix representation of the quantum gates used in the circuit~\cite{Markov_2009}. To give a brief description of the process, each matrix of a gate can be used as a building block to piece together the desired network. In graphical notation, the $2\times 2$ matrix of a single qubit gate is depicted by a single node with two adjoining edges. Likewise, the $4\times 4$ matrix of a two qubit gate is reshaped into a rank 4 tensor of dimension $2\times 2\times 2 \times 2$ and depicted by a single node with four adjoining edges. If each qubit in the circuit has an initial state given by a two dimensional complex vector (usually the ``zero'' state $\big(\begin{smallmatrix}1 \\ 0\end{smallmatrix}\big)$), the initial state of the $n$-qubit circuit is given by a network consisting of $n$ disjoint nodes with a single edge attached to each. Each qubit has a corresponding open index associated with it. To apply a gate to one or two of the qubits one needs only to adjoin the corresponding gate tensor to the network by attaching it to the open indices of the network corresponding to the target qubits. By adjoining gate tensors to the network in the order in which they appear in the quantum circuit, a tensor network representing the output state is constructed and should resemble a circuit diagram such as the one shown in Fig.~\ref{fig:QC_TN_expression}. The memory required to store the described tensor network structure grows linearly with the total number of qubits and gates in the circuit making it possible to represent states from large quantum circuits on a computer.

The price to be paid for such a memory efficient representation of the output state is a potentially large amount of computation to do whenever a probability amplitude is requested from the state. Namely, to retrieve a component from the output state, all tensors in the tensor network need to be multiplied together to compute the an expression analogous to equation~\eqref{tn_expression}. This computation is referred to as ``contracting'' the network and can be prohibitively expensive if not done carefully which we will elaborate on in the next section. Furthermore, contracting the network with $n$ open indices results in $2^n$ amplitudes being computed and stored in memory raising the same memory issue plaguing full wave function simulators. To this end, in order to compute a single amplitude, an additional layer of qubit tensors mirroring the tensors representing the initial state of the circuit, is adjoined to the network, closing all of its open indices. Assuming the final output tensors added to the network describe a state of the qubits given by some bitstring $x$, contracting the network will compute the single probability amplitude for $x$ occurring when the output of the circuit is measured.

To summarise, the main function of a tensor network simulator is to compute probability amplitudes for various possible measurement outcomes of a quantum circuit. This is done by using a description of the circuit to build a tensor network storing the output state of the circuit and then repeatedly contracting the network with different output tensors adjoined to it corresponding to the different outcomes.

\section{Scaling challenges}\label{sec:scaling_challenges}
When attempting to execute a large tensor network simulation of a quantum circuit on a distributed system, a number of challenges arise which we discuss in the subsections below. We loosely categorise these challenges as high and low level challenges to reflect when these challenges arise. High level challenges can be dealt with before the simulation begins (i.e. before any probability amplitudes are computed) while low level challenges occur during the simulation. We now describe these challenges and some of the solutions that can be used to overcome them.

\subsection{High-level challenges}
The main high level challenge for a tensor network simulator is to plan exactly how it will contract a network. Contracting a tensor network involves multiplying all of its tensors together which can always be done as a sequence of pairwise tensor contractions~\cite{Bridgeman_2017}, an operation which replaces two tensors in the network with their product. The computational cost of contracting a network typically depends on the sequence of pairwise contractions used to contract it. However, for a network consisting of $N$ tensors, there are $N!(N-1)!/2^{N-1}$ possible contraction orders and the majority of these can be very inefficient making it difficult to identify an optimal order. In fact, finding the optimal contraction order is equivalent to finding the tree decomposition of a graph with minimal treewidth~\cite{Markov_2009}, a problem which is believed to be NP-complete~\cite{fried_2018}. While finding the optimal order may be intractable, many algorithms exist to find ``good'' contraction orders in bounded time~\cite{fried_2018,Gray-2020}.

To determining efficient contraction orders, the method employed by the authors for the QuantEx simulator is described in \cite{Gray-2020} and is based on computing a tree decomposition with minimal treewidth for the network's line graph before converting it into a contraction order. In terms of tensor network contraction, the treewidth of a tree decomposition of a network's line graph is the rank of the largest intermediate tensor created when the network is contracted according to the order determined by the decomposition. Thus, minimising the treewidth is akin to minimising the largest intermediate tensor produced by the contraction process. An algorithm called FlowCutter~\cite{Hamann_2016} is used to construct tree decompositions with optimal treewidth of a graph by iteratively partitioning it using maximal flows on the graph.

\begin{figure}[!t]
\centering
\includegraphics[width=.5\textwidth]{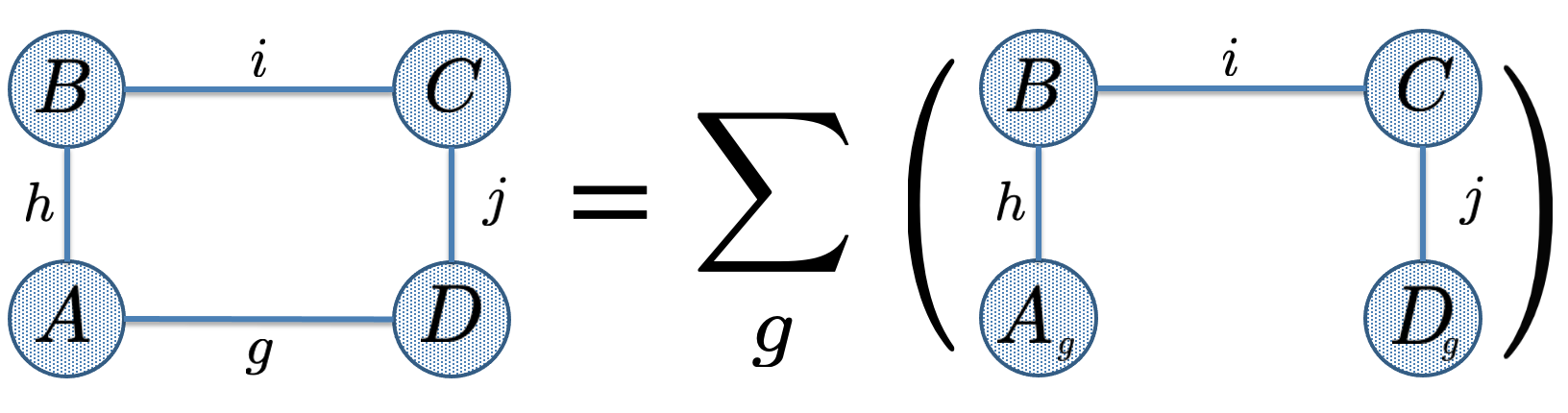}
\caption{Decomposing a tensor network as a sum of sliced networks.}
\label{fig:TN_slicing}
\end{figure}

Another high level challenge for a tensor network simulator is in deciding how to fit the network contraction into available memory resources. Even when an optimal contraction order is found for a network, the memory requirements for contracting a network may exceed the limits of the platform running the simulation due to large intermediate tensors being created during network contraction. A common method for tackling this issue is that of slicing~\cite{Huang_2021} which allows the simulator to reduce the memory requirements of a contraction in exchange for more network contractions. Slicing works by replacing the contraction of the original network with the contraction of several networks with less indices and summing the results. The contraction of the smaller networks are independent and can be done in parallel. The smaller networks are created from the original by choosing an index of the original network and fixing it to one of its values. In mathematical notation we can write this as follows:
\begin{equation}\label{tn_sliced_expression}
    \sum_{ghij}A_{gh}B_{hi}C_{ij}D_{gj} = \sum_g\big(\sum_{hijk}A_{gh}B_{hi}C_{ij}D_{gj}\big),
\end{equation}
where the expression in parentheses on the right hand side is that of a smaller tensor network as depicted in Fig.~\ref{fig:TN_slicing}. In order for this technique to adequately reduce the memory requirements of the simulation, the largest intermediate tensor produced by contracting the smaller networks must be smaller than the largest tensor created by contracting the original network. This may not be the case if indices are chosen to be sliced which are irrelevant to the size of the largest intermediate tensors. Thus, the ability to identify indices in the network which reduce the size of the largest intermediate tensors when sliced is integral to effectively using the method of slicing to reduce the memory requirements of a contraction. In graph theoretic terms, this problem translates into identifying edges of a graph which reduce the treewidth of the graph when removed, another NP-complete problem~\cite{shutski_2020}.

A novel algorithm for identifying efficient indices to slice in a tensor network was proposed in~\cite{shutski_2020} and was shown to perform well. Their greedy ``tree trimming'' algorithm always chooses to slice indices that reduce the size of the largest intermediate tensor and therefore the treewidth. When there are several largest tensors, it aims to choose an index that reduces the size of not just one of the largest tensors but also the greatest number of intermediate tensors. In the special case where there exist several such options, it chooses an index which reduces the overall memory resources of the network contraction the most, randomly breaking ties when no unique option exists.

\subsection{Low-level challenges}


The primary low level challenge of a tensor network simulator is to perform the basic operations constituting the task of contracting a network. For instance, while the challenge of effectively slicing a tensor network is classified as a high-level challenge, the operations required for its implementation may require efficient memory management and inter-node communication. Therefore, to facilitate tensor network slicing and contraction on a distributed system, it is necessary to implement tensor primitives that are capable of permitting a sliced network to be contracted in parallel, such as tensor slicing and recombination. Furthermore, tensor primitives for contracting, reshaping and permuting tensors are needed to perform a network contraction. With the emergence of Exascale systems, making efficient use of accelerators and CPU features is also crucial. While GPUs can offer impressive speedups, they also incur a second, tighter, memory bottleneck, adding further complexity to memory management.


As mentioned in Sec.~\ref{sec:overview}, to simulate a quantum circuit, random bitstrings need to be generated which are distributed according to the output state of the circuit. To achieve this, rejection sampling techniques can be employed. Optimising these methods can either be done by reducing the number of network contractions that need to be performed or by reducing the average cost of a network contraction~\cite{qflex_npj_2019, Pan_2021, Markov_2018}, raising yet another low level challenge. One issue with using a basic rejection sampling technique, or the frugal rejection method from~\cite{Markov_2018}, is that it requires setting a parameter $M$ which determines the average number of candidate probabilities computed before a candidate bitstring is accepted as an output bitstring. Setting $M$ to an optimal value requires prior knowledge of the simulated circuit's output distribution which may not be available for the user's circuit of interest. To avoid this issue, we chose to implement an alternative rejection method for QuantEx which was proposed in~\cite{Caffo_2002}. The empirical supremum rejection method follows the same recipe as the basic and frugal rejection methods but initially chooses $M=1$. Then, after each iteration of the algorithm, $M$ is updated according to 
\[
M = \mbox{max}(M,\ p(X)N),
\] 
where $X$ was the candidate bitstring in that iteration and $N$ is the number of possible bitstring candidates. The authors of~\cite{Caffo_2002} suggest that $M$ tends to converge quickly and suggest the use of a warm up period before using the algorithm, where several iterations of the algorithm are executed and their outputs discarded, to find a good estimate of the optimal $M$. A big advantage of this method is that it requires no prior knowledge of the circuit output distribution to find a value for $M$. This makes it ideal for sampling from a general quantum circuit.

\section{QuantEx}\label{sec:quantex}
\subsection{The QuantEx Tensor Network Simulator}
In this section, we introduce a tensor network based quantum circuit simulator referred to as QuantEx. It was developed by the authors and designed to be extensible, scalable on Exascale compute platforms and utilise the solutions to the high and low level challenges discussed in Sec.~\ref{sec:scaling_challenges}. The QuantEx framework consists of several special purpose software packages aiming to address different issues that arise in tensor network simulations. These are QXTools, QXTns, QXGraphDecompositions and QXContexts, each of which we described below and are available on github under the JuliaQX organisation\footnote{QuantEx Team, \url{https://https://github.com/JuliaQX}}. The packages are also registered in the Julia package registry making them easily accessible. Julia~\cite{bezanson_julia_2017} is used as the primary language, because of its flexible type system, the ability to wrap components in other languages while also providing native performance and native support for GPGPU programming. A domain specific language (DSL) is also used to represent a simulation as a set of primitive tensor operations. This separates the high level index accounting and contraction planning from the low level implementation of the tensor network operations and makes it easier to support new hardware and network architectures.

\textbf{QXTools} is the main QuantEx package for orchestrating a tensor network simulation of a quantum circuit. It can be used to create a tensor network for a quantum circuit, identify an efficient contraction scheme for the network and generate simulations files, including tensor data files and DSL files, that describe how the simulation should be executed on a cluster. It provides a quantum circuit simulation workflow which consists of the following steps:
\begin{enumerate}
    \item Circuits are built and represented as QXZoo circuits.
    
    \item The QXZoo circuit is converted to a QXTns tensor network.
    
    \item This network is converted to a graph data structure provided by QXGraphDecompositions and a suitable tree decomposition and set of edges to slice are identified.
    
    \item Using the tree decomposition and set of edges to slice a DSL representation of the computation is generated. This is then used as input to QXContexts to perform the computation using the context and settings that make the best use of the available resources.
\end{enumerate}

\textbf{QXZoo} Provides data structure and functions for representing and generating quantum circuits.

\textbf{QXTns} is a Julia package with data structures and utilities for manipulating tensor networks. As well as a generic tensor network data structure, it also contains specific data structures for handling tensor networks derived from quantum circuits.

\textbf{QXGraphDecompositions} is a package for analysing and manipulating graph structures describing tensor networks. It provides data structures and functions for analysing and manipulating graph representations of tensor networks. In particular, it provides functions for finding efficient tree decompositions and for identifying sets of indices which when sliced can reduce the treewidth of the selected tree decomposition. This makes it possible to distribute computations across multiple processes/nodes.

\textbf{QXContexts} is designed to parse the simulation files created by QXTools and perform the tensor contractions that constitute the circuit simulation making use of distributed compute resources via MPI as well as hardware accelerators. It provides implementations of the tensor primitives mentioned as one of the low level challenges in Sec.~\ref{sec:scaling_challenges} and uses the Julia package CUDA.jl \footnote{\url{https://github.com/JuliaGPU/CUDA.jl}} to provide NVIDIA GPU support. It also provides an implementation of the sampling algorithm discussed in the low-level challenges section of Section \ref{sec:scaling_challenges} which can be used to generate random bitstrings which are distributed according to the output state of the simulated quantum circuit.

Three levels of parallelism are used by QXContexts to decompose the computation. At the highest level, computing probability amplitudes for different bitstrings is embarrassingly parallel and is distributed across multiple processes using a MPI communicator. A second level of parallelism is available if slicing is used to avoid large memory requirements. A MPI sub-communicator is used to balance the subtasks created by slicing across processes. Finally, the lowest level of parallelism is in utilising GPU hardware to perform tensor contractions which can be mapped to matrix multiplications.

\subsection{Initial Performance Results}
Work is ongoing to procure performance results for QuantEx on PRACE tier-0 and pre-Exascale HPC systems and to optimise the framework. In preparation for this, smaller scale tests were conducted to test scaling, support for different CPU architectures and GPU acceleration. We present these results in this section. To evaluate the performance of the QuantEx software, we use as a test case the problem of computing probability amplitudes for a list of possible bitstring outputs of a quantum circuit. The quantum circuits we use in these test cases are instances of random quantum circuits (RQC) defined in~\cite{Boixo_2018} and used in Google’s quantum advantage experiments~\cite{Arute_2019}. These circuits consist of a 2 dimensional array of qubits with several layers of quantum gates acting on all qubits. For our initial scaling calculations, simulation files were generated for a RQC with a 5 by 5 grid of qubits and 24 layers of gates.

\begin{figure}[!t]
\centering
\includegraphics[width=.35\textwidth]{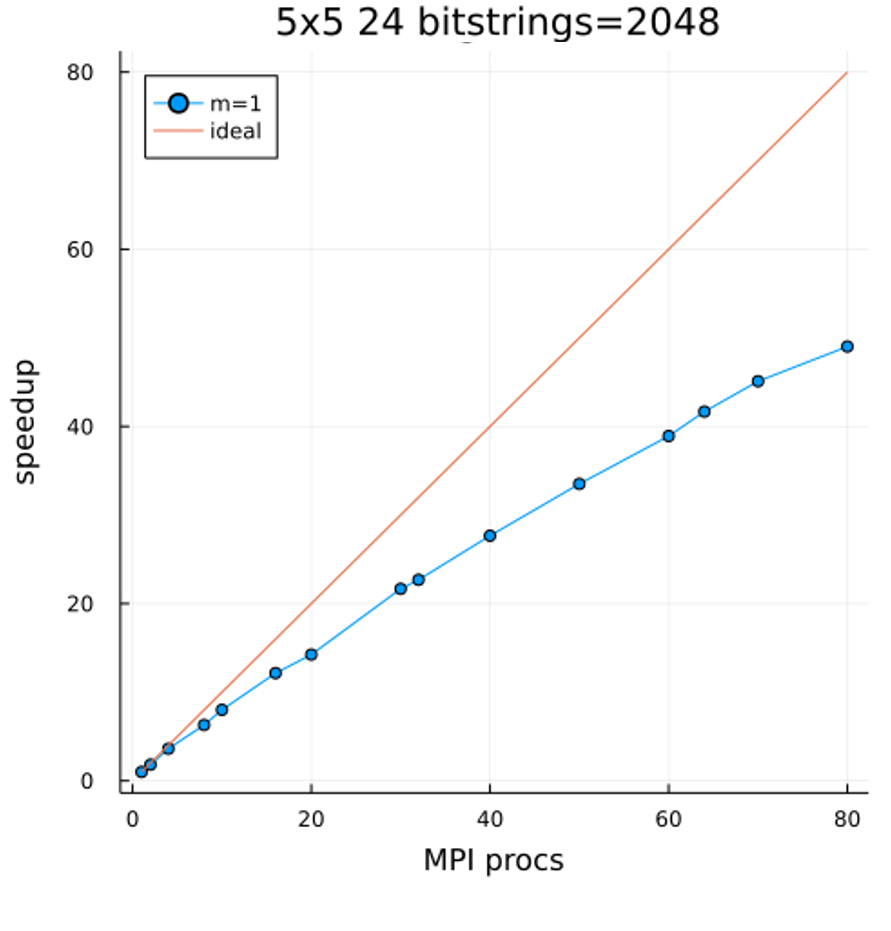}
\caption{Initial QuantEx scaling results on ICHEC's Kay. As a test case, 2048 amplitudes were computed for a RQC with a $5\times 5$ grid of qubits and 24 layers of gates and a single index was sliced.}
\label{fig:Kay_results}
\end{figure}

Initial scaling results were computed on ICHEC’s Kay cluster of 336 nodes where each node has 2x 20-core 2.4 GHz Intel Xeon Gold 6148 (Skylake) processors, 192 GiB of RAM, a 400 GiB local SSD for scratch space and a 100Gbit OmniPath network adaptor. For Fig. \ref{fig:Kay_results}, we take the case of computing 2048 amplitudes for the 5x5x24 RQC, with a single sliced bond, on 4 nodes with an increasing number of processes. The trend shown by these strong scaling results are expected to carry over to larger platforms given the embarrassingly parallel nature of how the simulation is decomposed.

Given that HPC systems are becoming increasingly heterogeneous, it is necessary for a viable simulator to run on various architectures in order to leverage novel HPC machines. To demonstrate that QuantEx is capable of this, additional tests of other HPC architectures have been performed on the BEAST system at LRZ. Small test quantum circuits consisting of 12 and 24 qubits were used in this case. The Bavarian Energy Architecture and Software Testbed (BEAST) is a collection of systems for the research and evaluation of new hardware technologies. Currently BEAST consists of three different CPU architectures: AMD X86, and Arm Fujitsu A64fx. An additional system segment is equipped with Arm ThunderX2, but the LIKWID tool was not fully functional on this architecture at the time of testing and therefore we leave it for future investigation. The AMD systems consists of two node Rome GPU 2U  servers, with two AMD EPYC 7742 with 64 cores along with 512GB of DDR4-3200, two 1.9 Terabyte SSD and two AMD Radeon MI-50 GPUs with 32 Gigabytes of high bandwidth memory (HBM). The interconnections between the nodes are Mellanox InfiniBand: HDR 200Gb/s. Finally, The Fujitsu A64fx system is an eight node HPE system consisting of Arm Fujitsu A64fx CPUs with 64 cores and two 512 bit vector units and 32 gigabytes HBM2 memory that is connected with a Mellanox InfiniBand EDR interconnect.

The table~\ref{tab:BEAST_results} lists the key features of the different architectures evaluated on BEAST. A point of reference (third column) is provided by the Intel Xeon Scalable Processors (``Skylake'') of SuperMUC-NG \footnote{\url{https://doku.lrz.de/display/PUBLIC/SuperMUC-NG}}. The table also presents some performance diagnostics collected using LIKWID, namely run time, arithmetic throughput, and measured memory bandwidth of our framework.

\begin{table}
\centering
\caption{\label{tab:BEAST_results}BEAST Results}
 \begin{tabular}{l || c | c | c} 
   & ARM   & AMD-ROME & Intel SKL \\
   & A64FX & EPYC 7742 & (SuperMUC-NG) \\
 \hline
 \hline
 Run-time (s)  & 0.14 & 0.030 & 0.061 \\
 12 Qubits & & & \\
 \hline
 Run-time (s) & 245.4 & 59.06 & 124.92 \\
 24 Qubits & & & \\
 \hline
 FLOPS DP & 93.74 & Not available & 196.26 \\
 (MFLOP/s) & & on this & \\
 12 Qubits & & architecture & \\
 \hline
 FLOPS DP & 823.17 & Not available & 511.7 \\
 (MFLOP/s) & & on this & \\
 24 Qubits & & architecture & \\
 \hline
 Memory Bandwidth & 389.44 & 496.4 & 546.99 \\
 (Mbytes/s) - 12 Qubits & & & \\
 \hline
 Memory Bandwidth & 1665.42 & 3110.72 & 4310.1 \\
 (Mbytes/s) - 24 Qubits & & & \\
 \hline
 Base-Frequency & 425 & 2250 & 2300 \\
 \hline
 SIMD (bit) & 2048 & 256 & 512 \\
 \hline
 Cores /node & 48 & 64 & 48 \\
 \hline
\end{tabular}
\end{table}

QuantEx’s NVIDIA GPU support was tested using a 16GB NVIDIA Volta V100 GPU on Cineca’s Marconi100 system\footnote{\url{https://www.hpc.cineca.it/hardware/marconi}}. Here, we measured the time to compute a single amplitude on both a GPU and on one of Marconi100’s 16-core IBM POWER9 processors. The time was measured for several, progressively difficult, quantum circuits and the results are displayed in table~\ref{tab:results}. Each row contains the results of a different circuit with the first column giving the name of the circuit. The second column contains the treewidth of the simulation which is a proxy for how difficult the simulation is as the complexity of the simulation is exponential in the treewidth~\cite{Markov_2009}. The memory column shows the maximum memory footprint of the process over the course of the calculation. The measured times show a clear benefit to using the GPU for larger circuits. Note, for the largest circuit (marked with asterisks in the Table), the memory footprint exceeded the memory capacity of the Volta V100. In this case, slicing was used to split the calculation into two subtasks and reduce the memory requirements of the calculation. The measured time is the time to complete one of these subtasks.


\begin{table}
\centering
\caption{\label{tab:results}Marconi100 Results}
 \begin{tabular}{c || c | c | c | c} 
 Circuit & Treewidth & Memory & CPU time & GPU time \\
 \hline
 \hline
 GHZ & 2 & 160B & 1.29ms & 2.538ms \\
 \hline
 RQC 4x4x24 & 11 & 18KB & 21.572ms & 53.372ms \\
 \hline
 RQC 6x6x24 & 19 & 4.2MB & 229.1ms & 112.1ms \\
 \hline
 RQC 5x5x32 & 23 & 67.1MB & 342.7ms & 167.9ms \\
 \hline
 RQC 7x7x24 & 24 & 128.5MB & 1.296s & 292.2ms \\
 \hline
 RQC 6x6x32 & 27 & 512MB & 9.141s & 384.9ms \\
 \hline
 RQC 7x7x32 & 32 & 16.4GB & 16s* & 1.25s* \\
 \hline
\end{tabular}
\end{table}

\section{Conclusion and Outlook}\label{sec:outlook}
We have provided a high level overview of the key challenges faced when scaling tensor network circuit simulation methods to Exascale and the some of the available solutions to those challenges. We also introduced the QuantEx simulation framework as a viable open source quantum circuit simulation tool and demonstrated it is capable of scaling on distributed systems and utilise GPU accelerators. The software was successfully tested on Intel and AMD CPUs and on NVIDIA GPUs and future work includes expanding the supported hardware to include AMD GPUs and Intel GPUs. While work is currently ongoing to benchmark and optimise the QuantEx simulator, it is the author's hope that the work presented here establishes QuantEx as an open source quantum circuit simulator with the potential to compete with state of the art simulators such as CoTenGra \footnote{\url{https://github.com/jcmgray/cotengra}} and QTensor \footnote{\url{https://github.com/Argonne-QIS/QTensor}} in the near future.

Future work on QuantEx the project includes further optimising circuit simulations and testing the software on the forthcoming european pre-Exascale and Exascale machines \cite{eurohpc_ju, german_exascale}. Optimizing simulations may be achieved via improved network contraction planning capabilities and better bitstring sampling methods. One method for improving a contraction plan for a tensor network is that of local optimization~\cite{Huang_2020}. This involves replacing subsections of a contraction plan with optimal alternatives found using an exhaustive search and potentially offers significant improvements in computational cost of a simulation. An approach to optimising bitstring sampling using memoization was also proposed recently~\cite{Pan_2021} and offers large reductions in the time complexity of a simulation. The method consists of designing a contraction plan for a tensor network with a natural checkpoint which a simulation can return to between samples to avoid recontracting a large portion of the network. A rewarding direction of future work would be to integrate this technique with QXContexts to greatly improve efficiency and possibly generalise the method to identify optimal checkpoints in arbitrary contraction plans, let alone carefully designed plans, broadening the contraction planning algorithms that can be used with this technique. 

Furthermore, efforts are ongoing to identify suitable opportunities to integrate the developed tools into commonly used quantum circuit simulation frameworks. One particular direction the QuantEx team is exploring is the possibility of integrating quantex as a backend for the popular Yao.jl \footnote{\url{https://github.com/QuantumBFS/Yao.jl}} framework. The Julia package YaoQX.jl \footnote{\url{github.com/JuliaQX/YaoQX.jl}} was developed with the hope of enabling Yao.jl users to take advantage of distributed systems and pre-Exascale and Exascale HPC clusters to simulate quantum circuits.

\section*{Acknowledgment}
This work was financially supported by the PRACE project funded in part by the EU’s Horizon 2020 Research and Innovation programme (2014-2020) under grant agreement 823767.

\vspace{12pt}

\end{document}